\documentstyle[12pt]{article}

\newcommand{\bm}[1]{\mbox{\boldmath $#1$}}
\newcommand{\be}{\begin{equation}}
\newcommand{\ee}{\end{equation}}
\newcommand{\ba}{\begin{eqnarray}}
\newcommand{\ea}{\end{eqnarray}}
\newcommand{\lb}{\label}
\newcommand{\ds}{\displaystyle}
\newcommand{\ra}{\rightarrow}
\newcommand{\ol}{\overline}

\newcommand{\bb}[1]{\bibitem{#1}}
\begin{document}
\begin{titlepage}
\setcounter{page}{1}
\title{Kaluza-Klein and Gauss-Bonnet\\
 cosmic strings}
\author{
\bigskip
Mustapha Azreg-Ainou$^1$
{\small and}
G\'erard Cl\'ement\thanks{E-mail: GECL@CCR.JUSSIEU.FR.}$\,\,^2$\\
\small $^1\,$ 5 rue du Soleil, 06100 Nice, France \\
\small $^2\,$ Laboratoire de Gravitation et Cosmologie Relativistes,\\
\small Universit\'e Pierre et Marie Curie, CNRS/URA769 \\
\small Tour 22-12, Bo\^{\i}te 142, 4 place Jussieu, 
75252 Paris cedex 05, France \\
}
\bigskip
\date{March 28, 1996}
\maketitle
\begin{abstract}
We make a systematic investigation of stationary cylindrically symmetric
solutions to the five-dimensional Einstein and Einstein-Gauss-Bonnet
equations. Apart from the five-dimensional neutral cosmic string metric,
we find two new exact solutions which qualify as cosmic strings, one
corresponding to an electrically charged cosmic string, the other to
an extended superconducting cosmic string surrounding a charged core. In
both cases, test particles are deflected away from the singular line
source. We extend both kinds of solutions to exact multi-cosmic string 
solutions.
\end{abstract}
\end{titlepage}

\setcounter{equation}{0}
\section{Introduction}
In four spacetime dimensions, the Einstein-Hilbert action augmented by a
cosmological constant term is the unique geometrical action (depending only
on the metric and the curvature tensor) leading to field equations which
involve at most second order derivatives of the metric. This is no longer
true if the dimensionality of spacetime is greater than four. As first
shown by Lovelock \cite{Love}, the most general second-order geometrical
action is a sum of terms proportional to the successive Euler forms, the
number of terms depending on the dimension. From a purely geometrical
perspective, there is no compelling reason to truncate this sum by keeping
only the Einstein-Hilbert action. The implications of these Lovelock
higher-dimensional theories of gravity have been explored in a number of
pioneering papers [2-8].

The simplest application of the Lovelock construction is to the case of
five dimensions. Kaluza and Klein \cite{KK} first suggested the
five-dimensional geometrical unification of gravity and electromagnetism,
both at the kinematical (equations of motion of charged particles) and 
dynamical (field equations) levels. The dynamical equations of the theory
\cite{JT} commonly known as Kaluza-Klein theory are simply the
five-dimensional Einstein equations. The dynamical equations of the more
general five-dimensional Lovelock theory are the Einstein-Gauss-Bonnet
equations (the single, quadratic extra term in the Lagrangian is the
Gauss-Bonnet \cite{pio} or Lanczos \cite{Madore85} invariant). As far as
we know, the only investigation of exact solutions to these
five-dimensional equations concerns a cosmological model \cite{DM86} (the
spherically symmetric solutions of \cite{Wheeler} are spherical in the
four space dimensions, and so do not have the Kaluza-Klein ${\rm V}_4
\times {\rm S}_1$ topology). One of the motivations for the present paper
was thus the search for exact non-trivial stationary solutions to the
Einstein-Gauss-Bonnet equations.

We shall focus in this paper on stationary cylindrically symmetric
spacetimes, with four commuting Killing vectors --- Kaluza-Klein cosmic
string solutions. Such solutions to the five-dimensional Einstein
equations were investigated by Ferrari \cite{Fer} in the static case.
Actually, Ferrari's assumption that the gravitational potential $g_{44}$
goes to a constant at spatial infinity restricted him to ultrastatic
solutions with $g_{44}$ constant; the corresponding Kasner-like four-space 
metrics are singular, except for the special case of the ``solenoid''
solution (see section 4 below). We shall relax these assumptions, and
enlarge our search to the Einstein-Gauss-Bonnet equations, with the hope
that singularities might be regularized by the Gauss-Bonnet term; this
will turn out to be true in the case of a particularly interesting
solution (section 6 below).

As shown by Witten \cite{Witten85}, some spontaneously broken gauge
theories lead to the appearance of superconducting cosmic strings carrying
a longitudinal electric current. The long-range behaviour of the
gravitational field of a straight superconducting cosmic string has been
derived by several authors \cite{SC} from the solution of the
Einstein-Maxwell equations with suitable extended or line sources. One
would expect to also be able to describe superconducting cosmic strings in
the five-dimensional unified theory, with possible differences with the
Einstein-Maxwell description due to the extra Kaluza-Klein scalar field
and to the Gauss-Bonnet non linear term. We shall show here that a certain
solution to the Einstein-Gauss-Bonnet equations with a line source may be
reinterpreted as a superconducting cosmic string solution to the
five-dimensional Einstein equations with an effective, Gauss-Bonnet
generated longitudinal electric current as extended source.    	

In the next section we write down in matrix form the stationary
cylindrically symmetric Einstein-Gauss-Bonnet equations. The third section
is devoted to an exhaustive classification of local solutions to the
five-dimensional Einstein equations (no Gauss-Bonnet coupling); the
generation of inequivalent global solutions from given local solutions is
briefly discussed in section 4. We next go over to the case of the full
Einstein-Gauss-Bonnet equations, and first investigate in section 5 a
special class of solutions, those for which both the Einstein and
Gauss-Bonnet contributions vanish independently; this class includes the
five-dimensional extension of the well-known neutral cosmic string
spacetime, as well as an electrically charged cosmic string solution. A
less trivial exact solution to the full Einstein-Gauss-Bonnet equations is
constructed in section 6, and interpreted as a superconducting cosmic
string. We then extend in section 7 both the charged and the
superconducting cosmic string solutions to exact multi-cosmic string
solutions. Our results are briefly summarized in the last section.

\setcounter{equation}{0}
\section{The Einstein-Gauss-Bonnet equations in the cylindrically
symmetric case}

 The most general geometrical action leading
to field equations containing at most second order partial derivatives is,
in the case of five spacetime dimensions \cite{Madore85},
\be
\lb{action}
S = -\frac{1}{8\,\pi\,G_{5}}\,\int d^5x\,\sqrt{g}\,
\left[ \alpha\, {\cal L}_0 + \frac{\beta}{2}\,{\cal L}_1 + \frac{\gamma}{4}\,
{\cal L}_2 \right], 
\ee
where ${\cal L}_0$, ${\cal L}_1$ and ${\cal L}_2$ are the cosmological,
Einstein-Hilbert and Gauss-Bonnet contributions, 
\be
{\cal L}_0 \equiv 1\,, \;\;\; {\cal L}_1 \equiv R\,,\;\;\;
{\cal L}_2 \equiv R^{ABCD}R_{ABCD} - 4\,R^{AB}R_{AB} + R^{2}
\ee
(upper-case Roman indices take the values 1 to 5). In this paper
we are mainly interested in classical solutions which approach cosmic string
metrics at large distances. This is possible only if we choose the
cosmological term to be absent, $\alpha = 0$. We also normalize the
five-dimensional gravitational constant $G_5$ by the choice $\beta = 1$.
The field equations deriving from the geometrical action (\ref{action}) 
are then
\be
\lb{eqs}
R_{AB} - \frac{1}{2}\,R\,g_{AB} + \gamma\,L_{AB} = 0,
\ee
where $L_{AB}$ is the covariantly conserved Lanczos tensor \cite{Madore85}
defined by
\be
\lb{L}
L_{AB} \equiv R_{A}\,^{CDE}R_{BCDE} - 2\,R^{CD}R_{ACBD} -
2\,R_{AC}R_{B}\,^{C} + RR_{AB} - \frac{1}{4}\,g_{AB}\,
{\cal L}_2\,.
\ee

A stationary cylindrically symmetric five-dimensional metric has four
commuting Killing vectors, one of which ($\partial_4$) is timelike, and two of
which ($\partial_2$ and $\partial_5$) have closed orbits. Choosing adapted 
coordinates
$x^1 = \rho$, $x^2 = \varphi$, $x^3 = z$, $x^4 = t$, $x^5$, such a metric
can be parametrized in terms of a $4 \times 4$ symmetrical matrix
$\lambda$ by
\be
\lb{met}
ds^{\,2} = - d{\rho}^{2} + {\lambda}_{ab}(\rho)\,dx^{a}\,dx^{b}
\ee
(with $a, b = 2, \ldots, 5$). A straightforward computation then gives
the Riemann tensor components
\ba
\lb{Riem1} & &R_{1ab1}=\frac{1}{2}\,(\lambda B)_{ab}\,, \\
\lb{Riem2} & &R_{abcd}=-\frac{1}{4}\,[(\lambda \chi)_{bc}\,(\lambda \chi)_{ad}-
(\lambda \chi)_{ac}\,(\lambda \chi)_{bd}],
\ea
in terms of the matrices
\be
\lb{chiB}
\chi \equiv {\lambda}^{-1}\,{\lambda}_{,\rho}\,,\;\;\;
B \equiv \chi_{,\rho} + \frac{1}{2}\,\chi^{2}\,.
\ee
From (\ref{Riem1}), (\ref{Riem2}) we compute the Ricci tensor components
\be \lb{Ric}
R_{11} = -\frac{1}{2}\,{\rm Tr}\,B\,,\;\;\; R^a\,_b = \frac{1}{2}
[ B - \frac{1}{2} \chi^2 + \frac{1}{2}({\rm Tr} \chi)\chi]\,^a\,_b \,,
\ee
as well as the Lanczos tensor components. We only give here as an example
one of the terms entering the computation of $L^a\,_b$,
\be
R^{acde}R_{bcde} = \frac{1}{8}\,[({\rm Tr}\chi^2)\chi^2 - \chi^4]^a\,_b\,.
\ee
We then arrive at the following form of the field equations (\ref{eqs}),
split into a scalar equation (the (11) components of (\ref{eqs})) and a
matrix equation, 
\ba
\lb{scal}
& 3\,\mbox{Tr}B+\frac{1}{2}\,[(\mbox{Tr}\chi)^{2}-\mbox{Tr}\chi^{2}]  
& + \gamma \left\{ \frac{1}{2}\,[\mbox{Tr}(B\chi^{2})
- \mbox{Tr}(B\chi)\,\mbox{Tr}\chi]\right. \nonumber \\
& & \left.+\frac{1}{4}\,\mbox{Tr}B\,[(\mbox{Tr}\chi)^{2}-\mbox{Tr}\chi^{2}]
\right\}=0\,,
\ea
\ba 
\lb{matr}
& & \chi_{,\rho}+2\,\mbox{Tr}\chi_{,\rho}+\frac{1}{2}\,[(\mbox{Tr}\chi)\chi+
\mbox{Tr}\chi^{2} +(\mbox{Tr}\chi)^{2}] + 
\gamma\,\left\{\frac{1}{2}\,(\chi^{3})_{,\rho}
-\frac{1}{2}\,(\mbox{Tr}\chi)(\chi^{2})_{,\rho}\right. \nonumber \\
& & \mbox{}+\frac{1}{2}\,[(\mbox{Tr}\chi)^{2}-\mbox{Tr}\chi^{2}]\,\chi_{,\rho}
- \frac{1}{2}\,(\mbox{Tr}\chi_{,\rho}) [\chi^{2}-(\mbox{Tr}\chi)\chi]
-\frac{1}{4}\,(\mbox{Tr}\chi^{2})_{,\rho}\,\chi \\
& & \left. \mbox{}+\frac{1}{4}\,[(\mbox{Tr}\chi)\chi^{3} - 
(\mbox{Tr}\chi^{2})\chi^{2} - (\mbox{Tr}\chi)  
(\mbox{Tr}\chi^{2})\chi + (\mbox{Tr}\chi)^{3}\chi] \right\} = 0\,. \nonumber
\ea

\section{Exact solutions of the Einstein equations}
\setcounter{equation}{0}
In this section we give the complete solution of equations (\ref{scal}),
(\ref{matr}) with $\gamma = 0$, i.e. of the stationary cylindrically
symmetric Kaluza-Klein equations. Let us define the four
invariants associated with the matrix $\chi$: 
\be
\lb{def}
f \equiv \mbox{Tr}\chi,\;\;\; g \equiv \mbox{Tr}\chi^{2},\;\;\;
h \equiv \mbox{Tr}\chi^{3},\;\;\; k \equiv \mbox{det}\chi\,.
\ee
Traces of higher powers of $\chi$ may be computed from these by using the
characteristic equation for the $4 \times 4$ matrix $\chi$,
\be
\lb{charac}
\chi^{4}-f\chi^{3}+\frac{1}{2}\,(f^{2}-g)\,\chi^{2}+\left(-\frac{1}{3}\,
h+\frac{1}{2}\,fg-\frac{1}{6}\,f^{3}\right)\chi+k \equiv 0\,.
\ee
For $\gamma = 0$ the field equations (\ref{scal}), (\ref{matr}) may be
written, using (\ref{def}), as
\ba
\lb{KK1}
& &3\,f_{,\rho}+\frac{1}{2}\,f^{2} + g = 0\,, \\
\lb{KK2}
& &\chi_{,\rho}+\frac{1}{2}\,f\chi +2\,f_{,\rho}+\frac{1}{2}\,f^{2}+
\frac{1}{2}\,g=0\,.
\ea
Tracing the matrix equation (\ref{KK2}), and eliminating $g$ between 
the resulting equation and (\ref{KK1}), we obtain the differential 
equation for $f$
\be
\lb{f}
f_{,\rho}+\frac{1}{2}\,f^{2}=0\,.
\ee
The invariant $g$ is determined from the solution of this equation by 
\be
\lb{g}
g = f^2\,.
\ee
The matrix $\chi$ may then be obtained by solving the linear equation
\be 
\lb{lin}
\chi_{,\rho} + \frac{1}{2}\,f\chi = 0\,.
\ee

The differential equation (\ref{f}) is solved either by
\be
f = \frac{2}{\rho}\,,
\ee
or by
\be
f = 0\,.
\ee
We first consider the case \underline{$f = 2/\rho$}. In this case, 
equation (\ref{lin}) integrates to 
\be
\lb{gen}
\chi = \frac{2}{\rho}\,A\,,
\ee
where $A$ is a constant real matrix constrained by
\be
\lb{constr}
\mbox{Tr}A^{2} = \mbox{Tr}A = 1\,.
\ee
Then, the integration of the first equation (\ref{chiB}) yields 
\be
\lb{solg}
\lambda = C\,\mbox{\large e}^{A\,\ln \rho^2}\,,
\ee
where $C$ is a constant real matrix of signature (-- -- + --). The symmetry
of the matrix $\lambda(\rho)$ is ensured by the symmetry conditions on C:
\be
\lb{symm}
C = C^{T}\,,\;\;\;CA = (CA)^T\,.
\ee

A matrix $A$ subject to the constraints (\ref{constr}) has four 
eigenvalues $p_i$ solving the characteristic equation
\be
p^{4}-p^{3}-\frac{1}{3}(c-1)p+d = 0\,,
\ee
where $c \equiv \mbox{Tr}A^{3}$, $d \equiv \mbox{det}A$. These eigenvalues
are related by the constraints 
\be
\lb{sum}
\sum_{i=1}^{4}p_{i} = 1, \qquad \sum_{i=1}^{4}p_{i}^{2} = 1\,.
\ee
The resulting metric is obviously of the Kasner type \cite{Kasner} if the 
four eigenvalues  $p_i$ are real. However, this by no means
exhausts the various solutions of the form (\ref{solg}). By considering
systematically the various matrix types for the matrix $A$, we arrive at
the following classification:

1) The four eigenvalues are complex,
\be
p_1 = x + iy\,,\;\; p_2 = x - iy\,,\;\; p_3 = \frac{1}{2} - x + iz\,,\;\; p_4 =
\frac{1}{2} - x - iz\,.
\ee
The Lagrange interpolation formula
\be
\lb{Lagrange}
f(A)=\sum_{i}f(p_{i})\prod_{j\neq i}\,\frac{A-p_{j}}{p_{i}-p_{j}}
\ee
then gives
\ba
\lb{sol1}
& & \lambda = \rho^{2x}\,CM\,[(A-x) \sin(2y \ln \rho + \delta_1) + y 
\cos(2y \ln \rho + \delta_1)] \nonumber \\
& & + \rho^{1-2x} CN\,[(A+x-\frac{1}{2}) \sin(2z \ln \rho + \delta_2) + z
\cos(2z \ln \rho + \delta_2)]\,,
\ea
where the matrices $M$ and $N$ are quadratic in $A$.

2) Two eigenvalues are complex and two real. The solution is obtained from
(\ref{sol1}) by the replacement $y \ra \ol{y} = iy$.

3) The four eigenvalues are real. The matrix $A$ may be diagonalized,
leading to the five-dimensional Kasner metric
\be
ds^{\,2} =
-d{\rho}^{2}-{\rho}^{2p_{1}}\,d{\varphi}^{\,2}-{\rho}^{2p_{2}}\,
dz^{\,2}+{\rho}^{2p_{3}}\,dt^{\,2}-{\rho}^{2p_{4}}\,
(dx^{5})^{\,2}\,,
\ee
where the $p_i$ are constrained by (\ref{sum})

4) Two of the real eigenvalues coincide. The Jordan normal form \cite{Gan}
of the matrix A is
\be
A=\left(\begin{array}{cccc} p_{1}&{\epsilon}_{1}&0&0\\0&p_{1}&0&0\\
0&0&p_{3}&0\\0&0&0&p_{4} \end{array} \right)
\ee
with $\epsilon_1 = 0$ or $1$. The resulting matrix $\lambda$ 
\be 
\lb{sol4}
\lambda = C\,\mbox{diag}({\rho}^{2p_{1}}, {\rho}^{2p_{1}},
{\rho}^{2p_{3}},
{\rho}^{2p_{4}})+{\epsilon}_{1}\,{\rho}^{2p_{1}}\,\ln{\rho}^{2}\,CH 
\ee
(with $(CH)_{ab} = 0$ except for $a = b = 2$) differs by a logarithm from
the Kasner metric.

5) The four eigenvalues are equal pairwise, $p_1 = p_2 = (1-\sqrt{3})/4,  
p_3 = p_4 = (1+\sqrt{3})/4$. The Jordan normal form of $A$ and the most
general associated matrix $C$ satisfying the symmetry conditions 
(\ref{symm}) are
\be 
A = \left( \begin{array}{cccc}
    p_{1}&{\epsilon}_{1}&0&0\\0&p_{1}&0&0\\ 
    0&0&p_{3}&{\epsilon}_{2}\\0&0&0&p_{3} \end{array} \right), \;\;
C = \left( \begin{array}{cccc}
    (1-{\epsilon}_{1})a_{1}&b_{1}&0&0\\b_{1}&c_{1}&0&0\\ 
    0&0&(1-{\epsilon}_{2})a_{2}&b_{2}\\0&0&b_{2}&c_{2}
\end{array} \right).
\ee
We then find that for $\epsilon_1 \epsilon_2 = 1$, $\det C \geq 0$ so that
the metric cannot have the Lorentz signature; for $\epsilon_1 \epsilon_2 =
0$, the solution is of the form (\ref{sol4}).

6) Three of the eigenvalues coincide and are nonzero, $p_1 = p_2 = p_3 = 
-p_4 = 1/2$. Putting $B \equiv A - 1/2$, with $B^4 = -B^3$, we obtain the 
solution
\be
\lambda = \rho C(1+B^{3})+2\rho\,\ln\rho\,C(B-B^{3})+
2\rho\,(\ln\rho)^{2}C(B^{2}+B^{3})-\frac{1}{\rho}\,CB^{3}\,,
\ee
with $B$ of the Jordan normal form
\be 
B = \left( \begin{array}{cccc}
    -1&0&0&0\\0&0&\epsilon_1&0\\0&0&0&\epsilon_2\\0&0&0&0
  \end{array} \right)\,.
\ee

7) Three of the eigenvalues vanish. Then the characteristic equation for A
reduces to
\be
A^4 = A^3\,,
\ee
leading to the solution
\be
\lb{sol7}
\lambda = C(1-A^{3})+2\ln{\rho}\,C\,(A-A^{3})+
2(\ln{\rho})^{2}\,C(A^{2}-A^{3})+{\rho}^{2}\,CA^{3}\,.
\ee
with $A$ of the Jordan normal form
\be
\lb{A7}
A = \left( \begin{array}{cccc}
           1&0&0&0\\0&0&\epsilon_1&0\\
           0&0&0&\epsilon_2\\0&0&0&0 \end{array} \right)\,.
\ee 
The solution (\ref{sol7}) with $\epsilon_1 \epsilon_2 = 1$ ($A^3 \neq
A^2$) is studied in Section 6. For $\epsilon_1 = 1$, $\epsilon_2 = 0$ ($A^3
= A^2 \neq A$) we recover a subcase of case 4 (see Section 5, eq. 
(\ref{charged})). For
$\epsilon_1 = \epsilon_2 = 0$ ($A^2 = A$), the solution (a particular
Kasner metric) is the product of a naked cosmic string metric by the Klein 
circle,
\be
\lb{neutr}
ds^{\,2}=-d{\rho}^{\,2}-{\alpha}^{2}{\rho}^{2}d{\varphi}^{\,2}-
dz^{\,2}+dt^{\,2}-(dx^{5})^{2}\,.
\ee

We now consider the second, trivial solution of the scalar equations
(\ref{f}) and (\ref{g}), \underline{$f = g = 0$}. Equation (\ref{lin}) 
is then solved by
\be
\lb{except}
\chi = A
\ee
where the constant matrix $A$ is now constrained by
\be
{\rm Tr}A = {\rm Tr}A^2 = 0\,.
\ee
The resulting metrical matrix $\lambda(\rho)$ is
\be
\lambda = C\,\mbox{\large e}^{A\,\rho}\,,
\ee
where the constant matrix $C$ again obeys the symmetry conditions
(\ref{symm}). From the characteristic equation 
\be 
\lb{charac0}
A^{4}-\frac{h}{3}\,A+k = 0\,,
\ee
it follows that if either $h$ or $k$ is nonzero, $A$ has either four
complex eigenvalues (the solution $\lambda(\rho)$ is then analogous to
(\ref{sol1}) with $\ln(\rho)$ replaced by $\rho$ \cite{these}) or two real
and two complex eigenvalues. We consider in more detail the special case 
$h = k = 0$, for which the characteristic equation (\ref{charac0}) reduces to
\be
A^4 = 0\,.
\ee
The solutions of this equation may be classified according to the rank of
the matrix $A$:

a) ${\rm r}(A) = 3$. The Jordan normal form of $A$ and the most general
associated matrix C are
\be
\lb{r3}
A =\left( \begin{array}{cccc}
          0&1&0&0\\0&0&1&0\\0&0&0&1\\0&0&0&0 \end{array} \right),\;\;
C =\left( \begin{array}{cccc}
          0&0&0&a\\0&0&a&b\\0&a&b&c\\a&b&c&d \end{array} \right).
\ee
It results from (\ref{r3}) that $\mbox{det}C \geq 0$, so that this 
solution does not lead to a Lorentzian metric.

b) ${\rm r}(A) = 2$. There are two possible Jordan normal forms for $A$.
The first form and a typical associated matrix $C$,
\be
A = \left( \begin{array}{cccc}
0&1&0&0\\0&0&1&0\\0&0&0&0\\0&0&0&0 \end{array} \right),\;\;\;
C = \left( \begin{array}{cccc}
0&0&-1&0\\0&-1&0&0\\-1&0&0&0\\0&0&0&-1 \end{array} \right),
\ee
lead to the five-dimensional metric
\be
ds^{\,2} = -\,d{\rho}^{\,2} - 2\,d\varphi\,dt 
- \frac{1}{2}\,(\rho\,dt+2\,dz)^2 + dz^2 - (dx^5)^2\,,
\ee
which is the product of a Petrov type N metric \cite{cyl} by the Klein
circle. In the case of the second normal form of $A$ with the most general
associated matrix $C$,
\be
A = \left( \begin{array}{cccc}
0&1&0&0\\0&0&0&0\\0&0&0&1\\0&0&0&0 \end{array} \right),\;\;\;
C = \left( \begin{array}{cccc}
0&a&0&b\\a&c&b&d\\0&b&0&e\\b&d&e&m \end{array} \right),
\ee
we find again ${\rm det}C \geq 0$ (non-Lorentzian metric).

c) ${\rm r}(A) = 1$. Then the Jordan normal matrix $A$ with a typical
matrix $C$
\be
A = \left( \begin{array}{cccc}
0&0&0&0\\0&0&1&0\\0&0&0&0\\0&0&0&0 \end{array} \right)\;,\;
C = \left( \begin{array}{cccc}
-1&0&0&0\\0&0&1&0\\0&1&0&0\\0&0&0&-1 \end{array} \right),
\ee
lead to the metric
\be
ds^{\,2} = -\,d \rho^{\,2} - \,d\varphi^{\,2} + 2\,dz\,dt + \rho\,dt^{\,2} 
-\,(dx^5)^{\,2}\,,
\ee
which is the product of a flat three-dimensional spacetime
\cite{Clement85} \cite{cyl} by two tori.

d) ${\rm r}(A) = 0$, i.e. $A = 0$. This obvious solution,
\be \lb{cyl}
ds^{\,2} = -\,d\rho^{\,2} -\,d\varphi^{\,2} -\,dz^{\,2 } +\,dt^{\,2} 
-\,(dx^5)^{\,2}\,,
\ee 
is the five-dimensional Minkowski metric with two dimensions ($\varphi, 
x^5$) compactified.

\section{From local to global solutions}
\setcounter{equation}{0}

The local solutions of the Kaluza-Klein equations given in the preceding
section are actually equivalence classes ---from each solution other
solutions may be obtained by linear coordinate transformations $x^a = 
L^a{}_b\,{x'}^b$ mixing the four commuting Killing vectors together (the
corresponding transformations on the matrices $A$ are similarity
transformations $A' = L^{-1}AL$). However, because two of our Killing
vectors ($\partial_2$ and $\partial_5$) have closed orbits, some of these
transformations (those with $L^a{}_m \neq \delta^a_m$ for $m = 2,5$) lead
to new solutions which are not globally equivalent to the old solutions.
Here we only give two examples of magnetic spacetimes obtained from
Minkowski spacetime by mixing $\partial_2$ and $\partial_ 5$ together.

In the first example (the Kaluza-Klein equivalent of the Melvin
\cite{Melvin} magnetic universe), the only non-zero 
non-diagonal element of the transformation matrix $L$ is $L^2{}_5 = \beta$, 
leading to the metric \cite{ident}
\be \lb{Fer1}
ds^2 =  - d\rho^2 - \rho^2\,(d\varphi + \beta\,dx^5)^2 - dz^2 + dt^2
- (dx^5)^2 \,,
\ee
where $\varphi$ is periodic with period $2 \pi$ and $x^5$ is periodic with
period $2 \pi a$.
Using the standard Kaluza-Klein decomposition of the five-dimensional metric 
\be \lb{dimred}
ds^2 = \ol{g}_{\mu \nu}\,dx^{\mu}\,dx^{\nu} + g_{55}\,(dx^5 +
kA_{\mu}dx^{\mu})^2 
\ee
($\mu, \nu = 1, \cdots, 4,$ and $k$ is the Kaluza constant) into  
four-scalar, four-vector and four-tensor fields $g_{55}$,
$A_{\mu}$ (electromagnetic potential), $\ol{g}_{\mu \nu}$
(four-dimensional metric), the metric (\ref{Fer1}) may be rewritten in
the form
\be \lb{Fer2}
ds^2 = - d\rho^2 - \frac{\rho^2}{1 + \beta^2 \rho^2}\,d\varphi^2 
- dz^2 + dt^2- (1 + \beta^2 \rho^2)[dx^5 + \frac{\beta \rho^2}{1 + \beta^2
\rho^2}\,d\varphi]^2
\ee
first given by Gibbons and Maeda \cite{GM}, and by Ferrari
\cite{Fer}. The resulting four-dimensional metric interpolates
between the Minkowski metric for $\rho = 0$ and the cylindrical Minkowski
metric (the four-dimensional part of (\ref{cyl})) for $\rho \ra \infty$, while
the azimuthal vector potential corresponds to a longitudinal magnetic field
$B^z = 2 k^{-1} \beta (1 + \beta^2 \rho^2)^{-3/2}$, and the scalar
field is parabolic. The magnetic flux through a plane $z =$ const. is 
\be
\Phi  = \oint_{\rho = \infty} A_{\mu}\,dx^{\mu} \;=\; \frac{2 \pi}{\beta k}\,. 
\ee

Our second example (flux string) is obtained from Minkowski spacetime by a
transformation matrix $L$ which is diagonal except for $L^5{}_2 = \nu$,
leading to the metric with a dislocation in the fifth dimension,
\be
ds^2 = - d\rho^2 - \rho^2\,d\varphi^2 - dz^2 + dt^2 - (dx^5 +
\nu\,d\varphi)^2 \,.
\ee
The azimuthal vector potential $A_{\varphi} = k^{-1}\nu$ is pure gauge,
except for a singularity on the axis $\rho = 0$. The corresponding 
longitudinal magnetic field $B^z = 2\pi k^{-1} \nu \,\delta^2 (x)$ 
is that of a flux tube concentrated on this axis. Other five-dimensional
generalizations of four-dimensional metrics singular on the axis $\rho =
0$ are generated by non-zero $L^2{}_2$ (the static cosmic string metric
(\ref{neutr})), $L^3{}_2$ (longitudinal dislocations \cite{dislo}), and
$L^4{}_2$ (spinning cosmic strings \cite{DJH}). 

\section{Solutions with vanishing Lanczos tensor}
\setcounter{equation}{0}

Now we consider the full five-dimensional Einstein-Gauss-Bonnet equations
(\ref{scal}), (\ref{matr}) with $\gamma \neq 0$. This system of coupled
first-order differential equations seems very difficult to disentangle, so
we first look for special solutions of the form
\be
\lb{special}
\chi = \alpha(\rho)\,A\,.
\ee
The $\gamma = 0$ solutions (\ref{gen}) and (\ref{except}) are of this
form. As shown in detail in \cite{these}, it turns out that the ansatz
(\ref{special}) selects the metrics for which the Einstein tensor vanishes
($\gamma = 0$ solutions) and the Lanczos tensor in (\ref{eqs}) also vanishes.

As we have seen in section 3, a first class of solutions of
the five-dimensional Einstein equations is given by
\be
\lb{an1}
\chi=\frac{2}{\rho}\,A\,, 
\ee
with ${\rm Tr}A = {\rm Tr}A^2 = 1$. Inserting (\ref{an1}) into the left-hand 
side of the full matrix equation (\ref{matr}), and expanding in powers
of $A$, we find that the term linear in $A$ vanishes identically, while
the combination of the terms in $A^2$ and $A^3$ reduces to
$-(8/\rho^4)\,(A^3-A^2)\,$. So our ansatz solves equation (\ref{matr}) for
$\gamma \neq 0$ only for matrices $A$ such that
\be
\lb{A3A2}
A^3 = A^2\,.
\ee
Computing the resulting matrix $B$ from (\ref{chiB}),
\be
B = \frac{2}{\rho}\,(A^2 - A)\,,
\ee
we then find ${\rm Tr}B = {\rm Tr}(B\chi) = {\rm Tr}(B\chi^2) = 0$, so
that the scalar equation (\ref{scal}) is also satisfied. Matrices $A$
obeying (\ref{A3A2}) are of the
Jordan normal form (\ref{A7}) with $\epsilon_1 \epsilon_2 = 0$. For
$\epsilon_1 = \epsilon_2 = 0$, the five-dimensional metric (\ref{neutr})
corresponds to a neutral cosmic string (or to a global, magnetic 
generalization such as (\ref{Fer1})). For $\epsilon_1 = 1$, $\epsilon_2
= 0$, we choose
\be
A=\left(\begin{array}{cccc} 1&0&0&0\\0&0&0&0\\0&0&0&1\\0&0&0&0
\end{array} \right),\;\;
C=\left(\begin{array}{cccc} -\alpha^2&0&0&0\\0&-1&0&0\\0&0&0&-1\\0&0&-1&0
\end{array} \right)
\ee
to obtain the five-dimensional metric
\be
\lb{charged}
ds^{\,2}=-d{\rho}^{\,2}-{\alpha}^{2}\,{\rho}^{2}\,d{\varphi}^{\,2}-dz^{\,2}
-2\,dt\,dx^{5} - 2\,\ln\rho\,(dx^{5})^{2}\,.
\ee

Following the standard Kaluza-Klein dimensional reduction (\ref{dimred}), 
we interpret the solution (\ref{charged}) as
describing a straight charged cosmic string with gravitational and
electric potentials 
\be
\ol{g}_{44} = \frac{1}{2\ln\rho} \,, \qquad
A_4 = \frac{1}{2k\ln\rho} \,.
\ee
There is however a problem with this interpretation, due to the fact that 
the scalar field $g_{55} = - 2\ln\rho$ does not go to a constant 
at spatial infinity. We recall that
the sourceless five-dimensional Einstein equations lead to the effective
four-dimensional Einstein equations with source,
\ba
& \ol{R}_{\mu}\,^{\nu} - \ds\frac{1}{2}\,\ol{R}\,\delta_{\mu}\,^{\nu} = &
\frac{k^2}{2}\, g_{55}\left( F_{\mu\rho}F^{\nu\rho} -
\frac{1}{4}\,\delta_{\mu}\,^{\nu} F_{\rho\sigma}F^{\rho\sigma} \right) 
\nonumber \\
& & + |g_{55}|^{-1/2}\,\ol{D}_{\mu}\ol{D}^{\nu}|g_{55}|^{1/2}\,.
\ea 
The source is the electromagnetic stress-energy tensor if $g_{55}(x)$
converges rapidly enough to a limiting value $g_{55}(\infty) = 
- 16\pi G/c^2k^2$. In the present case, $g_{55}(x)$ has no limit so that
the Kaluza-Klein constant $k$ is undetermined. This problem with infrared
logarithmic divergences is a familiar one in the case of infinite straight
string sources, and should disappear in the case of a closed charged
cosmic string. Geodesic motion in the five-dimensional metric
(\ref{charged}) is obtained by combining geodesic motion in the
four-dimensional metric (4.27) of \cite{cyl} (with $\omega^{\prime} = 0$, $a
= -1$) with a uniform motion in the $z$ direction. As discussed in
\cite{cyl}, the string singularity at $\rho = 0$ is ``harmless'', in the
sense that it deflects away all timelike and lightlike geodesics.
Another interesting property of the
metric (\ref{charged}) is that, for $\alpha = 1$, spatial sections are
flat, which suggests that multi-charged cosmic string solutions are also
possible. These shall be investigated in section 7.

We now turn to the second class of solutions of the five-dimensional
Einstein equations,
\be
\chi = A\,,
\ee
with ${\rm Tr}A = {\rm Tr}A^2 = 0$. Then the full equation (\ref{matr})
with $\gamma = 0$ is satisfied, while the full equation (\ref{scal}) gives
${\rm Tr}A^4 = 0$. This implies, from the trace of equation
(\ref{charac}), ${\rm det}A = 0$, so that the characteristic equation
reduces to a special case of equation (\ref{charac0}), 
\be
A^4 = p^3A\,,
\ee
showing that the matrix $A$ has the four eigenvalues $0$, $p$, $jp$,
$j^2p$ with $j = {\rm e}^{2i\pi/3}$. A typical solution is
\ba
\lb{A4A}
& ds^2 = & -d\rho^2 + 2 \cos\left( \frac{\sqrt{3}p\rho}{2} - \frac{\pi}{3} 
\right) e^{-p\rho/2}d\varphi^2 - 4 \cos\frac{\sqrt{3}p\rho}{2} 
{\rm e}^{-p\rho/2}d\varphi dt \nonumber \\ 
& & + 2 \cos \left( \frac{\sqrt{3}p\rho}{2} + \frac{\pi}{3} \right) 
 e^{-p\rho/2}dt^2 - {\rm e}^{p\rho}dz^2 - (dx^5)^2 \,. 
\ea
Such a metric, which does not approach, up to logarithms, a cosmic string
metric at large distances, does not admit a satisfactory physical
interpretation. Note that both solutions (\ref{charged}) and (\ref{A4A}) 
are the product
of the Klein circle or of the real line by a cylindrically symmetric
solution of the four-dimensional Einstein equations \cite{cyl}, which
explains why their Lanczos tensor vanishes identically \cite{Madore85}.

\section{A Gauss-Bonnet superconducting cosmic string}
\setcounter{equation}{0}

We now inquire whether there is a non-trivial ({\em i.e.} with non-zero
Lanczos tensor) solution of the five-dimensional Einstein-Gauss-Bonnet
equations, which at the same time is asymptotic, up to logarithms, to the
cosmic string metric (\ref{neutr}). Our strategy is to expand the matrix
$\lambda(\rho)$ in powers of $\gamma$,
\be
\lb{pert1}
\lambda(\rho) = \lambda_0(\rho) + \gamma \lambda_1(\rho) + \cdots \,,
\ee
where $\lambda_0(\rho)$ is an exact solution of the Einstein equations
($\gamma = 0$) with the desired asymptotic behaviour, and to solve
perturbatively the full $\gamma \neq 0$ equations to obtain the $n^{\rm th}$
order term $\lambda_n(\rho)$ from the lower order terms. Because the
Einstein-Gauss-Bonnet equations (\ref{scal}), (\ref{matr}) are non-linear equations in
$\chi = \lambda^{-1}\lambda_{,\rho}$, with the zeroth order $\chi_0(\rho)$
going as $1/\rho$ for the cosmic string metric (\ref{neutr}), the
asymptotic behaviour of the full solution (\ref{pert1}) will automatically
be governed by that of $\lambda_0(\rho)$. As an added bonus, we will find
that, owing to the special algebraic properties of $\chi_0$, the
linearized solution $\chi = \chi_0 + \gamma\chi_1$ is actually an exact
solution of the full non-linear equations!  

A solution of the five-dimensional Einstein equations which is asymptotic,
up to logarithms, to the cosmic string metric (\ref{neutr}) must be of the
form (\ref{solg}),
\be
\lambda_0 = C\,\mbox{\large e}^{A\,\ln \rho^2}\,,
\ee
where the matrix $A$ has the eigenvalues (1, 0, 0, 0), {\em i.e.} must
obey the algebraic relation
\be
\lb{alg}
A^{4} = A^{3}
\ee
without necessarily being diagonalizable. If $A^2 = A$, or $A^3 = A^2$
with $A^2 \neq A$, we recover the neutral cosmic string metric
(\ref{neutr}) or the charged cosmic string metric (\ref{charged})
respectively, which as we have seen solve trivially the
Einstein-Gauss-Bonnet equations. So we assume $A^3 \neq A^2$,
corresponding to the Jordan normal form (\ref{A7}) with $\epsilon_1 =
\epsilon_2 = 1$,
\be
\lb{Aex}
A=\left( \begin{array}{cccc}
         1&0&0&0\\0&0&1&0\\0&0&0&1\\0&0&0&0
\end {array} \right).
\ee 

Now we linearize the Einstein-Gauss-Bonnet equations (\ref{scal}), 
(\ref{matr}), using
the perturbative expansion for the matrix $\chi = \lambda^{-1}\lambda_{,\rho}$,
\be
\lb{pert2}
\chi(\rho) = \frac{2}{\rho}\,A + \gamma \chi_1(\rho) + \cdots \,.
\ee
The linearized equations are 
\ba
\lb{pert3}
& 3 {\rm Tr}\chi_{1,\rho} + {\ds\frac{1}{\rho}}[4{\rm Tr}(A\chi_1) & 
\!\!\! + \,\,2 {\rm Tr}\chi_1] = 0 \,, \nonumber \\
& \chi_{1,\rho} + 2 {\rm Tr}\chi_{1,\rho} + {\ds\frac{1}{\rho}} [\chi_1 & 
\!\!\! + \,\,({\rm Tr} \chi_1) A + 2{\rm Tr}(A\chi_1) + 2 {\rm Tr}\chi_1] \\
& &\!\!\! - \,\,{\ds\frac{8}{\rho^4}}(A^3 - A^2) = 0 \,.\nonumber
\ea
Combining the first equation with the trace of the second equation, we
obtain the two scalar equations
\be
{\rm Tr}(A\chi_1) = {\rm Tr}\chi_1\,, \qquad
{\rm Tr}\chi_ {1,\rho} + \frac{2}{\rho}{\rm Tr}\chi_1 = 0 \,.
\ee
The second of these equations is solved by
\be
{\rm Tr}\chi_1 = \frac{2c}{\rho^2}
\ee
($c$ constant). The second equation (\ref{pert3}) now reduces to
\be
\chi_{1,\rho} + \frac{1}{\rho}\chi_1 + \frac{2c}{\rho^3}A -
\frac{8}{\rho^4}(A^3 - A^2) = 0 \,,
\ee
which is solved by
\be
\chi_1 = \frac{2}{\rho}D + \frac{2c}{\rho^2}A - \frac{4}{\rho^3}(A^3 -
A^2) \,,
\ee
with $D$ a constant matrix such that ${\rm Tr}(AD) = {\rm Tr}D = 0$.
Actually the first two terms result from the expansion to first order of
$\chi = 2(A + \gamma D)/(\rho - \gamma c)$, and may be gauged away by
suitable coordinate transformations. So the genuine first-order
perturbation is
\be
\chi_1 = - \frac{4}{\rho^3}(A^3-A^2) \,.
\ee
It is now very easy to check, using the properties $A \chi_1 = \chi_1 A =
0$, $\chi_1\,^2 = 0$, ${\rm Tr}\chi_1 = 0$, that
\be
\lb{exact1}
\chi=\frac{2}{\rho}\,A-\frac{4\,\gamma}{{\rho}^{3}}\,(A^{3}-A^{2})
\ee
is an exact solution of the full Einstein-Gauss-Bonnet equations 
(\ref{scal}), (\ref{matr}).

The corresponding metrical matrix, obtained by solving the differential
equation $\lambda_{,\rho} = \lambda \chi$, is
\be
\lambda = C \exp{[A \ln \rho^2 + \frac{2\gamma}{\rho^2}(A^3 - A^2)]} =
C[{\rm e}^{\ds A \ln \rho^2} + \frac{2\gamma}{\rho^2}(A^3 - A^2)] \,.
\ee
Choosing $A$ in the Jordan normal form (\ref{Aex}) and
\be
C=\left( \begin{array}{cccc}
         -\alpha^2&0&0&0\\0&0&0&-1\\0&0&-1&a\\0&-1&a&b
\end {array} \right)\,,
\ee 
we finally obtain the exact solution
\ba
\lb{exact2}
& ds^{\,2} & = - d{\rho}^{\,2} - {\alpha}^{2}\,{\rho}^{2}\,d{\varphi}^{\,2}
-2\,dz\,dx^{5} - dt^{\,2} - 4L\,dt\,dx^5 \nonumber \\
& & -2(L^2 - p - \gamma/\rho^2)\,(dx^5)^2 \,, \qquad 
L = \ln(\rho/\rho_0) \,,
\ea
where we have defined $\ln\rho_0 = a/2$, $p = (a^2 + 2b)/4$. The
corresponding
four-dimensional metric and electromagnetic potentials are, according to
the Kaluza-Klein dimensional reduction (\ref{dimred}),
\ba
\lb{exact3}
\ol{g}_{\mu\nu}\,dx^{\mu}dx^{\nu} & = & - d\rho^2 - \alpha^2 \rho^2
\,d\varphi^2 - \frac{1}{2(L^2 + p + \gamma/\rho^2)}\,dz^2 \nonumber \\
& & + \frac{(L^2 + p + \gamma/\rho^2)}{(L^2 - p - \gamma/\rho^2)}
\left( dt + \frac{L}{(L^2 + p + \gamma/\rho^2)}\,dz \right)^2 \!, \\
A_{\mu}\,dx^{\mu} & = & \frac{1}{2k(L^2 - p - \gamma/\rho^2)}\
(dz + 2L\,dt) \,. \nonumber
\ea
As in the case of the charged cosmic string solution (\ref{charged}), the
scale of the electromagnetic field, proportional to the inverse of the
Kaluza-Klein constant $k$, is undetermined because the scalar field
$g_{55}(x)$ does not go to a constant at spatial infinity. However we
expect the cylindrical solution to approximate the behaviour of the
gravitational and electromagnetic fields not too far from a closed line
source modelled by $\rho = 0$.

The electromagnetic field (\ref{exact3}) is mainly electric, with a small
magnetic component, at both large ($\rho \gg \rho_0$) and small
($\rho \ll \rho_0$) distances. This suggests that the distributional
source for this metric is a line of electric charge, as borne out by a
careful computation of the left-hand side of the Gauss-Bonnet equations
(\ref{eqs}). However, we shall argue that these equations may be
reinterpreted as the five-dimensional equations
\be
\lb{eff}
{R^A}_B-\frac{1}{2}\,R\,{\delta^A}_B = 8\pi G\,
T_{\mbox{\tiny eff}\,B}^{A}
\ee
with an effective source $T_{\mbox{\tiny eff}\,B}^{A}$ which is the sum 
of a distributional contribution and of the continuous Gauss-Bonnet 
contribution $-(\gamma/8\pi G){L^A}_B$. The computation of the left-hand
side of (\ref{eff}) involves the Ricci tensor components given, from
equations (\ref{Ric}) and (\ref{exact1}), by
\be
{R^a}_b = \frac{1}{2\rho}\,(\rho \chi)_{,\,\rho} = {[\,A\,\Delta \ln \rho +
\gamma\,(A^3 - A^2)\,\Delta \rho^{-2}\,]^a}_b \,,
\ee
where $\Delta$ is the covariant Laplacian. Going over to two-dimensional
conformal coordinates ($x$, $y$) with
\be
(x + iy)^{\alpha} = \rho\,{\rm e}^{i \alpha \varphi}\,,
\ee
and using $\Delta\ln\rho = 2\,\pi\,\alpha\,
{\delta}^{2}(\bm{x})$, where ${\delta}^{2}(\bm{x})$ is the two-dimensional
covariant Dirac distribution, we obtain from (\ref{Aex}) the non-vanishing
components of the effective source,
\ba
& &T_{\mbox{\tiny eff}\,3}^{3}=T_{\mbox{\tiny eff}\,4}^{4}=
T_{\mbox{\tiny eff}\,5}^{5}=\frac{1-\alpha}{4G}\,
{\delta}^{2}(\bm{x})\,, \nonumber \\ 
& &T_{\mbox{\tiny eff}\,4}^{3}=T_{\mbox{\tiny eff}\,5}^{4}=\frac{\alpha}{4G}
\,{\delta}^{2}(\bm{x})\,,\\
& &T_{\mbox{\tiny eff}\,5}^{3}=-\frac{\gamma}{2\pi G}\,\frac{1}{{\rho}^{4}} 
\,. \nonumber
\ea
Recalling that the energy-momentum tensor components 
$T_{\mbox{\tiny eff}\,5}^{\mu}$ are, in the Kaluza-Klein theory,
proportional to the electromagnetic current density $j^{\mu}$, we are thus
led to interpret the metric (\ref{exact2}) as describing an extended
superconducting cosmic string (longitudinal current density 
$T_{\mbox{\tiny eff}\,5}^{3}$) surrounding a longitudinally boosted
electrically charged naked cosmic string. 

Another effect of the Gauss-Bonnet coupling is to transform the mild
logarithmic singularity of the metric (\ref{exact2}) for $\rho \ra 0$ into
a strong $\rho^{-2}$ singularity, which at first sight is rather annoying.
However, as we shall see by studying geodesic motion in this metric, while
the singularity mildly repels test particles for $\gamma = 0$, it becomes
strongly repulsive for $\gamma > 0$, so that in this sense the $\gamma >
0$ solution is less singular than the $\gamma = 0$ one. The first
integrated five-dimensional geodesic equations in a stationary 
cylindrically symmetric metric (\ref{met}) are
\ba 
\lb{geo}
& &\left(\frac{d\rho}{d\tau}\right)^{2}-{\Pi}_{a}\,{\lambda}^{ab}(\rho)\,
{\Pi}_{b}+\epsilon=0 \,, \\
& &\frac{dx^a}{d\tau\,}={\lambda}^{ab}(\rho)\,{\Pi}_{b}\,,
\ea
where $\tau$ is an affine parameter, $\epsilon$ is a real constant, and
the $\Pi_a$ are the constants of the motion associated with the four
Killing vectors. In the case of the metric (\ref{exact2}), equation
(\ref{geo}) reads (after rescaling lengths so that $\rho_0 = 1$)
\ba \lb{geo2}
& \left(\ds \frac{d\rho}{d\tau}\right)^2 + (\alpha^{-2} \Pi_2^2 + 2 \gamma
\Pi_3^2)\, \rho^{-2} + 2 \Pi_3^2\, (\ln \rho)^2 - 4 \Pi_3 \Pi_4 \ln \rho 
& \nonumber \\
&  + 2 \Pi_3 (p\,\Pi_3 + \Pi_5) + \Pi_4^2 + \epsilon = 0 \,. & 
\ea	 
For $\gamma < 0, \,\Pi_3^2 > - \Pi_2^2/2 \gamma \alpha^2$, the effective
potential is strongly attractive, and all geodesics terminate at the
singularity $\rho = 0$. For $\gamma = 0$, the effective potential is
repulsive except if $\Pi_2 = \Pi_3 = 0$, allowing only spacelike geodesics
($\epsilon < 0$) to reach the singularity. For $\gamma > 0$, the
centrifugal repulsion is enhanced by a term proportional to $\Pi_3^2$, so
that again only spacelike geodesics can reach the singularity. We also
note that the asymptotic behaviour of the effective potential is dominated
by the term $2 \Pi_3^2 (\ln \rho)^2$ so that, whatever the value of the
Gauss-Bonnet coupling constant $\gamma$, only geodesics with $\Pi_3 = 0$
and $\epsilon < 0$ extend to infinity; the fact that timelike or lightlike
geodesics are bounded is again a pathology of infinite straight cosmic 
strings.

\setcounter{equation}{0}
\section{Multiple cosmic strings}
In this section we show how the non-Kasner solutions (\ref{charged}) and
(\ref{exact2}) may be extended to multi-cosmic string solutions, using the
construction of \cite{PLA}, which we first briefly review. Multi-cosmic
string spacetimes having less symmetry than the one-cosmic string
spacetimes consided so far, we only assume here the existence of two
commuting Killing vectors, one of which is $\partial_5$, and the other is
either $\partial_t$ (stationary configurations) or $\partial_z$ (parallel
cosmic strings). In this case the five-dimensional metric can be
parametrized by
\be \lb{met2}
ds^2 = \tau^{-1}h_{ij}\,dx^i\,dx^j +
\mu_{ab}(\,dx^a + A_i^a\,dx^i)(\,dx^b + A_j^b\,dx^j)\,,
\ee
where $i$ takes the three values (1, 2, 3) or (1, 2, 4), $a$ takes the two
values (4, 5) or (3, 5), and $\tau = - {\rm det}\,\mu_{ab}$. Define the
twist two-vector $\omega_a$ such that
\be \lb{twist}
\omega_{a,i} = h^{-1/2} \tau \mu_{ab} h_{il} \epsilon^{jkl}
A_{j,k}^b 
\ee
($h = - {\rm det}h_{ij}$, and $\epsilon^{jkl}$ is the antisymmetric symbol),
and the $3 \times 3$ matrix field
\be
M = \left( \begin{array}{cc}
\mu_{ab} + \tau^{-1} \omega_a \omega_b & - \tau^{-1} \omega_a \\
- \tau^{-1} \omega_b & \tau^{-1}
\end {array} \right).
\ee   
The five-dimensional Einstein equations then reduce to the
three-dimensional sigma-model system \cite{Maison}
\be \lb{Maison}
(M^{-1}M^{,i})_{;i} = 0 \,, \qquad R_{ij} = \frac{1}{4} \, {\rm Tr}
(M^{-1}M_{,i}M^{-1}M_{,j}) \,,
\ee
where all geometric symbols refer to the three-dimensional metric $h_{ij}$.
If this metric is (pseudo-)Euclidean, then a class of solutions to the
equations (\ref{Maison}) is given by
\be \lb{multi}
M = C\,{\rm e}^{\ds \,A \sigma(\bm{x})}\,{\rm e}^{\ds\,A^2 \phi(\bm{x})} \,,
\ee
where the $3 \times 3$ matrices $A$ and $C$ are constrained by 
\be
{\rm Tr} A = {\rm Tr}A^2 = 0\,, \qquad C = C^T\,, \qquad CA = (CA)^T \,,
\ee
and $\sigma(\bm{x})$, $\phi(\bm{x})$ are arbitrary harmonic functions,
\be \lb{harm}
\nabla^2 \sigma = 0\,, \qquad \nabla^2 \phi = 0 \,,
\ee
where $\nabla^2$ is the flat-space Laplacian. The linearity of equations
(\ref{harm}) then leads to the existence of multi-centre \cite{PLA} or, in
the present case, of multi-cosmic string solutions to the original field 
equations.
 
The solution (\ref{charged}) with $\alpha = 1$ may be put in the form
(\ref{met2}) with
\be \lb{muchar}
\mu = \left( \begin{array}{cc}
0 & -1 \\
-1 & -2 \ln \rho
\end{array} \right)
\ee     
($\tau = 1$), $A_i^a = 0$, and $h_{ij} = - \delta_{ij}$ ($i = 1, 2, 3$).
The matrix $M$ associated with (\ref{muchar}) is of the form (\ref{multi})
with 
\be
A = \left( \begin{array}{ccc}
0 & 1 & 0 \\
0 & 0 & 0 \\
0 & 0 & 0
\end{array} \right), \qquad 
C = \left( \begin{array}{ccc}
0 & -1 & 0 \\
-1 & 0 & 0 \\
0 & 0 & 1
\end{array} \right)
\ee
($A^2 = 0$), and $\sigma(\bm{x}) = 2 \ln \rho$, where $\rho$ is the
distance of the point $\bm{x}$ to the charged cosmic string. The natural
generalization to a system of non-parallel charged cosmic strings is the
linear superposition of harmonic functions
\be
\sigma(\bm{x}) = 2 \sum_{\alpha} c_{\alpha}\,\ln \rho_{\alpha} \,,
\ee
where $\rho_{\alpha}$ is the distance of the point $\bm{x}$ to the 
$\alpha^{\rm th}$
cosmic string, and $c_{\alpha}$ is an arbitrary weight. The resulting
multi-cosmic string metric is 
\be
ds^2 = - dx^2 - dy^2 - dz^2 - 2 \,dt\,dx^5 - \sigma(\bm{x})\,(dx^5)^2\,.
\ee
Owing to the special algebraic character of this metric, the only
non-vanishing Riemann tensor component for this spacetime is $R_{5ij5} =
-(1/2) \sigma_{,i,j}$, and the resulting Lanczos tensor vanishes
identically, as in the one-cosmic string case (the harmonicity condition
on $\sigma$ is not used to prove this result, which also holds on the
charged cosmic string sources).

The construction is more involved in the case of the solution
(\ref{exact2}) with $\alpha = 1$, $\gamma = 0$, which may be rewritten in
the form (\ref{met2}) with $i = (1, 2, 4)$, $a = (3, 5)$. Explicitly,
\be \lb{exact4}
ds^2 = - dx^2 - dy^2 - dt^2 +
\mu_{ab}(\,dx^a + A_i^a\,dx^i)(\,dx^b + A_j^b\,dx^j)\,,
\ee
with
\be
\mu = \left( \begin{array}{cc}
0 & -1 \\
-1 & -2(L^2 - p)
\end{array} \right), \qquad
A_i^a = 2L\,\delta_i^4\,\delta_3^a 
\ee       
($\tau = 1$), and $L = \ln \rho$ (taking $\rho_0$ as the unit of length).
The corresponding twist field
is, according to (\ref{twist}), $\omega_3 = 0$, $\omega_5 = 2 \varphi$.
The resulting matrix $M$ may be put in the form (\ref{multi}) with
\be
A = \left( \begin{array}{ccc}
0 & 0 & -1 \\
0 & 0 & 0 \\
0 & 1 & 0
\end{array} \right), \qquad 
C = \left( \begin{array}{ccc}
0 & -1 & 0 \\
-1 & 2p & 0 \\
0 & 0 & 1
\end{array} \right)
\ee
($A^3 = 0$), and
\be \lb{harm2}
\sigma = - 2\,{\cal I}m \ln \zeta \,, \qquad \phi = -2\,{\cal R}e \, 
(\ln \zeta)^2\,,
\ee
with $\zeta(\bm{x}) = \rho\,{\rm e}^{i \varphi}$. This may be
generalized to a multi-cosmic string solution of the five-dimensional
Einstein equations by replacing (\ref{harm2}) by the linear superpositions
\ba \lb{harm3}
& \sigma = - 2\,{\cal I}m \sum_{\alpha} c_{\alpha} \ln \zeta_{\alpha} \,, 
\qquad \phi = -2\,{\cal R}e \, 
\sum_{\alpha} c_{\alpha}^2 (\ln \zeta_{\alpha})^2 \,, & \nonumber \\
& \ln \zeta_{\alpha} = \ln\rho_{\alpha} + i\,\varphi_{\alpha} &
\ea
of harmonic functions singular on cosmic strings rotated and translated in
three-dimensional Euclidean space $(x, y, t)$ with respect to the original
cosmic string, $\rho_{\alpha}$ being the Euclidean distance of the event 
$\bm{x} = (x, y, t)$ to the $\alpha^{\rm th}$ cosmic string, and 
$\varphi_{\alpha}$ its azimuthal angle relative to this string. From the
corresponding matrix $M$, we extract $\mu_{ab}$ and $\omega_a$, and invert
equation (\ref{twist}) which reduces in this case to the linear equation
\be
(\nabla \wedge A^3)_{\,i} = \sigma_{,\,i} \,,
\ee
to obtain the metric generated by a system of parallel cosmic strings,
moving relative to each other with uniform velocities. This metric is of the
form (\ref{exact4}) with
\be \lb{exact5}
\mu = \left( \begin{array}{cc}
0 & -1 \\
-1 & - \sigma^2/2 + \phi + 2p
\end{array} \right), \qquad
\begin{array}{ll}
A_i^3 = 2\,\sum_{\alpha}
c_{\alpha}\,\hat{u}_{\alpha}^i\,\ln \rho_{\alpha} \\
A_i^5 = 0 
\end{array} ,
\ee       
where $\hat{u}_{\alpha}$ is the direction in $(x, y, t)$ space of the
$\alpha^{th}$ cosmic string. 

Finally we extend our multi-cosmic string solution to the case of the
Einstein-Gauss-Bonnet equations with $\gamma \neq 0$. We again make the
ansatz (\ref{exact5}) with the same harmonic function $\sigma$ as in
(\ref{harm3}), but a modified function $\phi$. Noting that the only
non-vanishing Riemann tensor components for this metric are
\be
R_{ijk5} = \frac{1}{2} \,\epsilon_{ijl}\,\sigma_{,\,k,\,l} \,, \;\;\; 
R_{i5j5} = \frac{1}{2} (\sigma \sigma_{,\,i,\,j} - \phi_{,\,i,\,j}) +
\frac{1}{4} \,(3 \sigma_{,\,i} \sigma_{,\,j} - \delta_{ij}
(\sigma_{,\,k})^2) \,,
\ee
we find that the non-vanishing Lanczos tensor component 
$L_{55} = - (1/2) (\sigma_{,\,k,\,l})^2$ is balanced by the Ricci tensor
contribution $R_{55} = (1/2) \nabla^2 \phi$ for the choice
\be
\phi = \phi_0 + \frac{\gamma}{2}(\nabla \sigma)^2
\ee
(where $\phi_0$ is the harmonic contribution in (\ref{harm3})),
corresponding to an exact multi-superconducting cosmic string solution of
the full Einstein-Gauss-Bonnet equations.

\section{Conclusion}
Apart from the trivial five-dimensional extension of the four-dimensional
neutral cosmic string solution of the vacuum Einstein equations and its
global extensions, we have obtained essentially two new exact cosmic
string solutions to the five-dimensional Einstein-Gauss-Bonnet equations. 

The electrically charged cosmic string metric (\ref{charged}) is a solution
independently of the value of the Gauss-Bonnet coupling constant $\gamma$. 
Its line source is not ``seen'' by massive or massless test particles, which 
are always deflected away without encountering the singularity. On the other
hand, the form of the solution (\ref{exact2}) depends explicitly on the value 
of $\gamma$. By reinterpreting the Gauss-Bonnet term as an effective source
term, we have identified this five-dimensional metric as that of an
extended superconducting cosmic string surrounding a charged core. Again,
for $\gamma \geq 0$ test particles are deflected away from the singular
line source. We have also shown that both solutions may be extended to
exact multi-cosmic string solutions, describing static systems of
non-parallel charged cosmic strings in the first case, systems of parallel
superconducting cosmic strings in uniform relative motion in the second case.

We have not investigated the stability of these classical solutions. The
behaviour of test particles leads us to conjecture that the charged cosmic
string solution, as well as the superconducting cosmic string solution for
$\gamma \geq 0$, which both bind test particles, are stable, while the
superconducting cosmic string solution for $\gamma < 0$ is unstable
through collapse onto the line singularity. 

\bigskip
\bigskip
\bigskip
{\Large \bf Acknowledgments}

\bigskip
One of us (M. A.-A.) wishes to thank M. Le Bellac for the kind hospitality 
afforded in his group at the Institut Non Lin\'eaire de Nice, where part 
of this work was done. We acknowledge stimulating discussions with B. Linet 
and J. Madore.

\end{document}